\documentclass{article}
\usepackage[utf8]{inputenc}

\usepackage{tikz}
\usetikzlibrary{positioning}
\usetikzlibrary{shapes.geometric}
\usetikzlibrary{shapes.misc}

\usepackage{amsfonts}
\usepackage{hyperref}
\usepackage{todonotes}
\usepackage{xcolor}
\usepackage{listings}

\definecolor{mGreen}{rgb}{0,0.6,0}
\definecolor{mGray}{rgb}{0.5,0.5,0.5}
\definecolor{mPurple}{rgb}{0.58,0,0.82}
\definecolor{backgroundColour}{rgb}{0.95,0.95,0.92}

\lstdefinestyle{CStyle}{
    backgroundcolor=\color{backgroundColour},   
    commentstyle=\color{mGreen},
    keywordstyle=\color{magenta},
    numberstyle=\tiny\color{mGray},
    stringstyle=\color{mPurple},
    basicstyle=\small\ttfamily,
    breakatwhitespace=false,         
    breaklines=true,                 
    captionpos=b,                    
    keepspaces=true,                 
    numbers=left,                    
    numbersep=-5 pt,                  
    showspaces=false,                
    showstringspaces=false,
    showtabs=false,                  
    tabsize=2,
    language=C
}

\newtheorem{example}{Example}

\title{A Protocol for Emotions}
\author{Gabriele Costa}
\date{IMT School for Advanced Studies\\
Lucca, Italy\\
\url{gabriele.costa@imtlucca.it}}

\begin{document}

\maketitle

\begin{abstract}
We tend to consider emotions a manifestation of our innermost nature of human beings.
Emotions characterize our lives in many ways and they chaperon every rational activity we carry out.
Despite their pervasiveness, there are still many things we ignore about emotions.
Among them, our understanding of how living beings transfer emotions is limited.
In particular, there are highly sophisticated interactions between human beings that we would like to comprehend.
For instance, think of a movie director who knows in advance the strong emotional impact that a certain scene will have on the spectators.
Although many artists rely on some emotional devices, their talent and vision are still the key factors.

In this work we analyze high-level protocols for transferring emotions between two intelligent agents.
To the best of our knowledge, this is the first attempt to use communication protocols for modeling the exchange of human emotions.
By means of a number of examples, we show that our protocols adequately model the engagement of the two parties.
Beyond the theoretical interest, our proposal can provide a stepping stone for several applications that we also discuss in this paper.
\end{abstract}

\section{Introduction}

Understanding the fundamental mechanisms that rule the exchange of emotional information between living beings has always fascinated scientists, philosophers and artists.
There are many reasons behind this interest, a deeper comprehension of how human mind works is possibly one of the most suggestive.
As a matter of fact, in human communications the emotional and informative contents are strongly interconnected.
For instance, consider the sentence ``this building is burning''.
A computer program may understand this piece of information and, for instance, trigger a fire alarm. 
Nevertheless, it will not feel in danger for this sentence.
On the other hand, it is very unlikely that a human hearing the previous sentence can detach the information from the emotional content.
As also suggested by~\cite{Wheater04emotion}, in human beings knowledge and emotions cannot be fully isolated.

It is reasonable that emotions have emerged in biological life forms for granting some evolutionary advantage and Darwin already noticed this in 1872~\cite{Darwin72emotions}.
For instance, some emotions are useful for self-diagnosis purposes, e.g., anxiety and disgust, while others may help in combining a large number of complex stimuli, e.g., fear or joy.
However, the reason why we communicate our emotions is less clear.
In some cases, emotions are triggered instinctively and the communication is based on some sort of automatism.
For instance, a ferocious animal showing its teeth is probably trying to instill fear in an opponent.
In some cases emotions can spread among a population of individuals as an epidemic.
As a consequence, we often say that things like laugh or panic are contagious.

More interesting, there are emotional transfers that are even more complex and require high level abstraction capabilities.
One example is empathy, i.e., the ability of perceiving and understanding the emotional state of others, even when this is not intentionally communicated.
On the opposite, humans can even intentionally induce in others emotional states that are unrelated to their own.
This is, for instance, the case in various types of artwork.
A stand up comedian can result hilarious to the public, independently from her own feelings and, even more, we can enjoy jokes and horror stories even when reading them in a book.
Often we also recognize the talent and intelligence of an artist by the ability to load her work with strong emotional contents.

In this paper we propose a new model for describing the emotional communications.
The basic idea is that, similarly to any other piece of information, emotions are transferred through protocols.
In particular, we consider a specific class, called \emph{$\Sigma$-protocols}, having a fundamental role in the theory of communication protocols. 
The reason why $\Sigma$-protocols look promising is manifold.
Primarily, they represent the building block for Proof-of-Knowledge (PoK), i.e., the process through which a verifier checks the truth of a claim made by someone, called the prover.
Despite their generality, these protocols are very compact (only 3 messages), which makes them adequate to implement efficient communications.
The role of efficiency also appears non negligible and, in general, timing is crucial for stimulating the desired emotions.

The main contributions of this paper are the following.

\begin{enumerate}
    \item A new theory of emotional communications, based on a variant of $\Sigma$-protocols that we call \emph{aporia protocols}.
    \item A detailed characterization of the main building blocks of aporia protocols, including common knowledge, timing, incongruity and distances.
    \item A discussion of implications and possible applications, including a new version of the classical Turing Test, an algorithm for processing emotions, an approach for detecting emotional communications in other life forms and an encoding of emotional states in computer interactions.
\end{enumerate}

All the propositions on which we base the current proposal are demonstrated through examples taken from various sources and artworks in general.
Although aporia protocols might not be a universal model of emotional interactions between intelligent beings, they appear to be general enough to capture many real cases of interest.
In this respect, this new model may shed light on some problems that have been debated for centuries.

The rest of this paper is structured as follows. In Section~\ref{sec:background} we briefly recall some background topics. Section~\ref{sec:aporia} presents our proposal for aporia protocols. In Section~\ref{sec:poe} we discuss the proof of emotion test and its applications. Section~\ref{sec:related} surveys on the related work and Section~\ref{sec:conclusion} concludes the paper.

\section{Background}
\label{sec:background}

In this section we briefly recall some background notions that are relevant for the correct understanding of this work.

\subsection{Proof of Knowledge}

Proof of Knowledge (PoK)~\cite{Bellare99pok} refers to a family of protocols that involve two parties, called \emph{Prover} and \emph{Verifier}.
The goal of the protocol is to allow the Verifier to ascertain that Prover knows a private piece of information $w$.
PoK includes a large variety of protocols of interest such as zero-knowledge proof and fair exchange protocols.
Although PoK protocols can be implemented in many ways, \emph{$\Sigma$-protocols}~\cite{Cramer94sigma} represent a standard approach.
The reason is both theoretical, since every $\Sigma$-protocol is also a PoK, and practical, due to their compact, three-message structure.

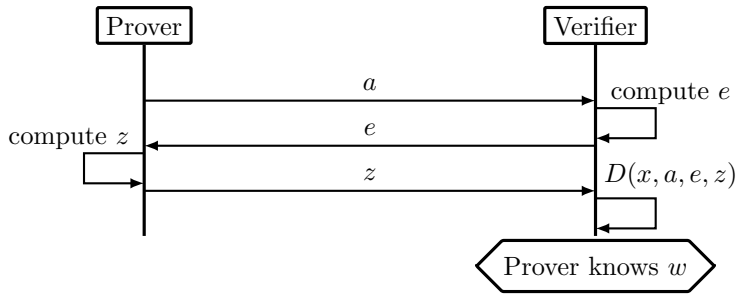
\begin{figure}[t]
  \centering
  
  \begin{tikzpicture}[every node/.append style={very thick,rounded corners=0.2mm}]
    
    \node[draw,rectangle] (Prover) at (0,0) {Prover};
    
    \node[draw,rectangle] (Verifier) at (6,0) {Verifier};
    
    \draw [very thick] (Verifier)--++(0,-2.8);
    \draw [very thick] (Prover)--++(0,-2.8);
    
    \draw [-latex,thick] (0,-1)--node [auto] {$a$}++(6,0);
    
    \draw [latex-,thick] (0,-1.6)--node [auto] {$e$}++(6,0);
    
    \draw [-latex,thick] (0,-2.2)--node [auto] {$z$}++(6,0);
    
    \node at (7,-0.9) {compute $e$};
    \draw[-latex,thick] (6,-1.1) -- +(0.8,0) |- (6,-1.5);
    
    \node at (-1,-1.5) {compute $z$};
    \draw[-latex,thick] (0,-1.7) -- +(-0.8,0) |- (0,-2.1);
    
    \node at (7,-2) {$D(x, a, e, z)$};
    \draw[-latex,thick] (6,-2.3) -- +(0.8,0) |- (6,-2.7);
    
    \node[chamfered rectangle, chamfered rectangle xsep=2cm, draw] (prop) at (6,-3.2) {Prover knows $w$};
    
    \end{tikzpicture}

    \caption{Abstract structure of $\Sigma$-protocols.}
    \label{fig:sigma}
\end{figure}

The abstract structure of a $\Sigma$-protocol is given in Figure~\ref{fig:sigma}.
Initially, Prover and Verifier share a common knowledge $x$.
Moreover, Prover knows a private datum $w$ such that $x R w$, for some relationship $R$.
Prover begins the protocol by sending a setup message $a$.
Then, Verifier computes and sends a challenge $e$.
Intuitively, the challenge is an input for a task that Prover cannot\footnote{Efficiently or with a non-negligible probability.} solve without knowing $w$.
To win the challenge, Prover computes a response $z$ which is then transmitted and validated by Verifier.
The validation amounts to a decision function $D$ which may involve every piece of information known by Verifier, i.e., $x$, $a$, $e$ and $z$.
If the validation succeeds, the protocol terminates with Verifier ascertaining Prover's knowledge of $w$.

\begin{example}
Consider the scenario in which Elliot claims to have control over the bank account of Ron.
Ron checks his balance and answers with the challenge ``withdraw $n$\$'' (where $n$ is a random amount, e.g., within $1$ and $100$).
After a few seconds Elliot answers ``Done''.
If Ron finds his balance reduced by exactly $n$\$ he accepts the claim of Elliot.
\end{example}

It is worth noticing that, in the previous example, Elliot is not disclosing the access mechanism $w$.
For instance, he might know Ron's credentials or he might have violated the bank service.
An alternative protocol in which Ron's challenge is that Elliot reveals one digit of his pin would not enjoy this property.  

\subsection{Aporia and emotions}

Many philosophers have considered human emotions in their works.
Among them, some authors discussed the strong relationship between emotional states and aporia.
For instance, the following definition of aporia appears in~\cite{Candiotto15aporia}.

\begin{quote}
The aporia is a mental state of perplexity and being at a loss, that
involves feelings, which in turn play a role in the cognitive development of the interlocutor. 
The aporetic state is not a purely cognitive state; it is a cognitively motivational state involving emotive elements.
\end{quote}

Typically, we perceive a certain level of aporia when observing two or more subjects that must be put in correlation, but do not fit well together.
Classical examples are illusions and puns, where two alternative, mutually exclusive interpretations of the same input are possible, e.g., as in Rubin's vase.
Also, a sense of aporia can arise when the comparison between two measurable inputs is difficult or illusory.
This happens, for instance, in illusions such as the checker shadow~\cite{Adelson00lightness} and the thermal grill~\cite{Craig94thermal}. 
However, aporia can also occur in other cases, e.g., for circular and recursive inputs.
Some prominent examples are the liar paradox and ``who came first, the chicken or the egg?''.

Many authors agree that a strong relationship exist between aporia and emotions.
The influential ``incongruity theory'' for humor is based on this assumption.
One can easily observe that most, possibly every joke use aporia in some way.
Nevertheless, aporia alone is not enough to stimulate specific emotional states.
When aporia comes in isolation from other communication devices, as in the case of optical illusions, often it is perceived as emotionally neutral.
This suggests that aporia is necessary, but not sufficient to induce humorous effects as well as other emotions (perhaps apart for mere stupor).

\subsection{Turing Test}

\begin{figure}[t]
    \centering
    \begin{tabular}{c | c}
    \includegraphics[height=0.2\textheight]{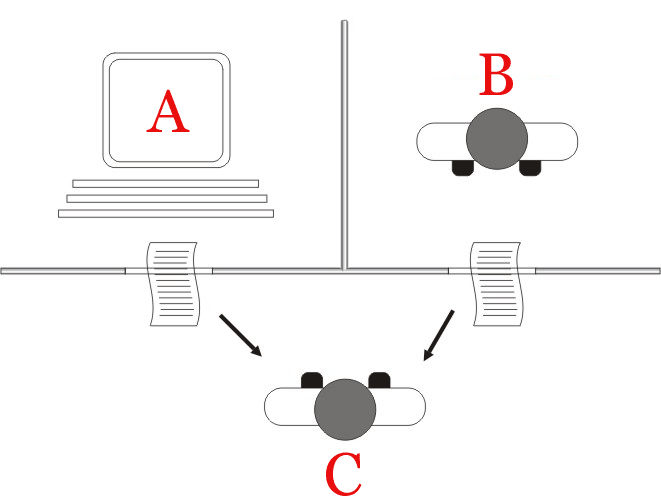} &
    \includegraphics[height=0.2\textheight]{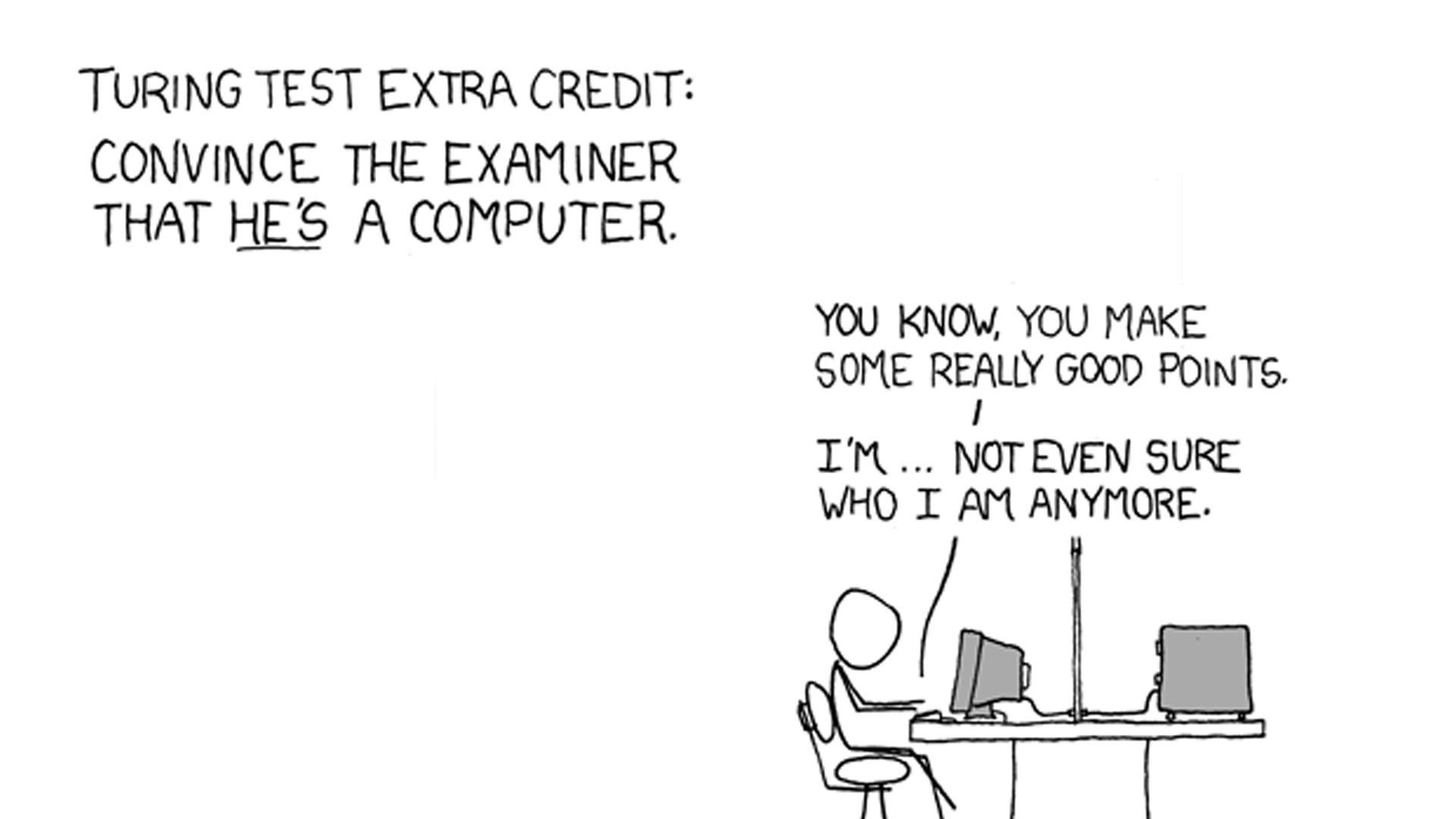}
    \end{tabular}
    \caption{The Turing Test (source: Wikipedia) and an XKCD cartoon on it.}
    \label{fig:tt}
\end{figure}

Historically, the Turing Test (TT) has been considered the touchstone of artificial thinking.
Briefly, TT is based on observational equivalence: if a machine (A) cannot be distinguished from a human (B) by an interviewer (C), the machine should be considered truly intelligent. 
Figure~\ref{fig:tt} shows the classical interpretation of TT (left) and a XKCD cartoon (right) underlying the fairness issue (see below).

In the last decades, many debated on the adequacy of TT for measuring the intelligence of a computer program.\footnote{We refer the interested reader to~\cite{Saygin00turing} for a survey.}
The main reason behind most criticism is the lack of agreement about the interviewer, i.e., which communication channels and which evaluation criteria should be used.
For instance, some authors claim that the interviewer should be allowed to use any type of sensory stimuli~\cite{Harnad91ttt} or even resort to biochemical analysis~\cite{Harnad02tttt}.
These objections have to do with the \emph{fairness} of TT, e.g., following~\cite{Harnad91ttt} a deaf could fail the test.
In this work we consider a fair version of TT, where the two participants must initially agree on a common language, i.e., which channels and symbols they will use to communicate during the test.
Both parts can retreat from the test if an agreement is not found and, in that case, the test is considered inconclusive.
Also, in case the test is carried out correctly, the interviewer can take into account the used language when taking the final decision.

\section{From Aporia to Emotions}
\label{sec:aporia}

In this section we present our proposal for protocols of emotions.
We start by introducing the general structure of protocols for stimulating aporia in a listener.

\subsection{Aporia protocol}

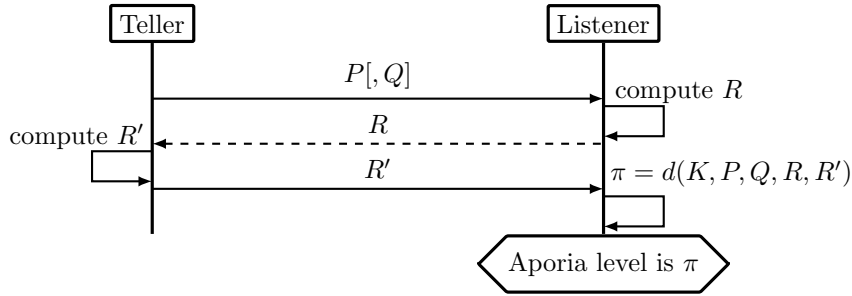
\begin{figure}[t]
  \centering
  
  \begin{tikzpicture}[every node/.append style={very thick,rounded corners=0.2mm}]
    
    \node[draw,rectangle] (Prover) at (0,0) {Teller};
    
    \node[draw,rectangle] (Verifier) at (6,0) {Listener};
    
    \draw [very thick] (Verifier)--++(0,-2.8);
    \draw [very thick] (Prover)--++(0,-2.8);
    
    \draw [-latex,thick] (0,-1)--node [auto] {$P[,Q]$}++(6,0);
    
    \draw [latex-,thick,dashed] (0,-1.6)--node [auto] {$R$}++(6,0);
    
    \draw [-latex,thick] (0,-2.2)--node [auto] {$R'$}++(6,0);
    
    \node at (7,-0.9) {compute $R$};
    \draw[-latex,thick] (6,-1.1) -- +(0.8,0) |- (6,-1.5);
    
    \node at (-1,-1.5) {compute $R'$};
    \draw[-latex,thick] (0,-1.7) -- +(-0.8,0) |- (0,-2.1);
    
    \node at (7.7,-2) {$\pi = d(K, P, Q, R, R')$};
    \draw[-latex,thick] (6,-2.3) -- +(0.8,0) |- (6,-2.7);
    
    \node[chamfered rectangle, chamfered rectangle xsep=2cm, draw] (prop) at (6,-3.2) {Aporia level is $\pi$};
    
    \end{tikzpicture}

    \caption{Abstract structure of the aporia protocol.}
    \label{fig:aporia}
\end{figure}

Intuitively, the protocol involves a Teller and a Listener.
The goal of the Teller is to induce a certain level of aporia in the Listener.
The protocol follows the $\Sigma$-protocol structure and it is given in Figure~\ref{fig:aporia}.
Initially, Teller and Listener share a common knowledge $K$.
For instance, $K$ can be seen as the context or the cultural background of Teller and Listener.
Teller starts the protocol by sending the pair $P,Q$, where $P$ is a premise of some sort, e.g., an anecdote, and $Q$ is a question about $P$, e.g., ``what is going to happen after $P$?''.
In some cases, $Q$ is not stated explicitly, e.g., when it is clear from the context.
Then, Listener computes its answer $R$ to $Q$ and sends it back to the Teller.
Again, answer $R$ can be implicit, e.g., if Teller can guess it.
Subsequently, Teller sends $R'$, i.e., an alternative answer for $Q$, possibly after computing it.
Eventually, Listener applies a function $d$ to measure the aporia level $\pi$ between $R$ and $R'$.
Also, when computing $\pi$, Listener can take into account any piece of information between $K$, $P$ and $Q$.

\begin{example}
The toilet dinner is possibly the most famous scene from ``The phantom of liberty'' (1974), by the celebrated Spanish director Luis Bu\~nuel.
The protocol between the scene/director (Teller) and the spectator (Listener) can be modeled as follows.
\begin{itemize}
\item[$K$.] Social conventions shared by most citizens of western countries.
\item[$P$.] At the beginning of the scene, a married couple is welcome by the household and his wife. They carry out formal pleasantries, introducing their young daughter and a family friend. Household's wife says they are ready to start.
\item[$Q$.] (implicit) What is happening?
\item[$R$.] (implicit) They are going to dine convivially.
\item[$R'$.] They defecate convivially.
\item[$\pi$.] Is (subjectively) the discrepancy between $R$ and $R'$ as continuations of $P$ under social conventions in $K$.
\end{itemize}
\end{example}

Although high levels of aporia can provide the stimulus to trigger an emotional reaction in the Listener, it is not enough by itself.
In general, high aporia level without an associated emotional state are infrequent and difficult to obtain.
This state of mind may result in alienation and, not surprisingly, it has been of interest for many authors, e.g., think of Samuel Beckett's ``Waiting for Godot''.\footnote{There, a prominent example is provided by Pozzo's sudden change, from wealthy aristocrat to disgraced, blind man.}

\subsection{Tone and emotion}

In general, the mechanism triggering a specific emotion in the Listener, may be complex and involve several aspects.
For instance, previous experiences and subjective belief may influence the interpretation of $P$.
Nevertheless, it is reasonable to assume that the overall tone used by the Teller during the protocol plays a central role. 
If we consider a neutral Listener, the main emotional stimuli might be those provided by the Teller, e.g., think of the background music during a movie scene.
To better highlight this phenomenon, consider the following example.

\begin{example}
The ``Scary Movie''\footnote{This name was inspired by the title of Kevin Williamson's script that became ``Scream'' afterwards.} saga consists of five movies, all written and directed by the Wayans brothers.
Each movie is a parody of one or more, mainly horror, films.
Parodies are obtained replicating the same structure of the original horror scenes, with only minimal changes, mostly related to the final punch line.
Thus, in terms of aporia, they often share the same structure (see Figure~\ref{fig:scary}).
For instance, consider the following protocol for Scream.

\begin{figure}[t]
    \centering
    \includegraphics[height=0.18\textheight]{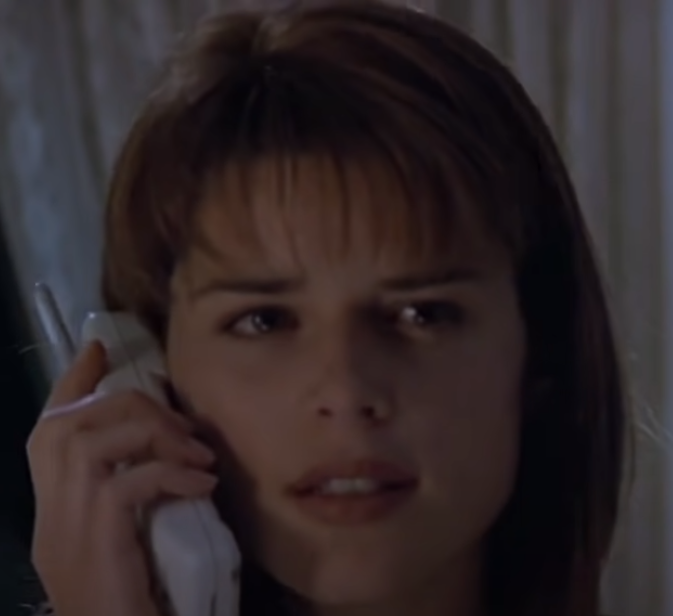}
    \includegraphics[height=0.18\textheight]{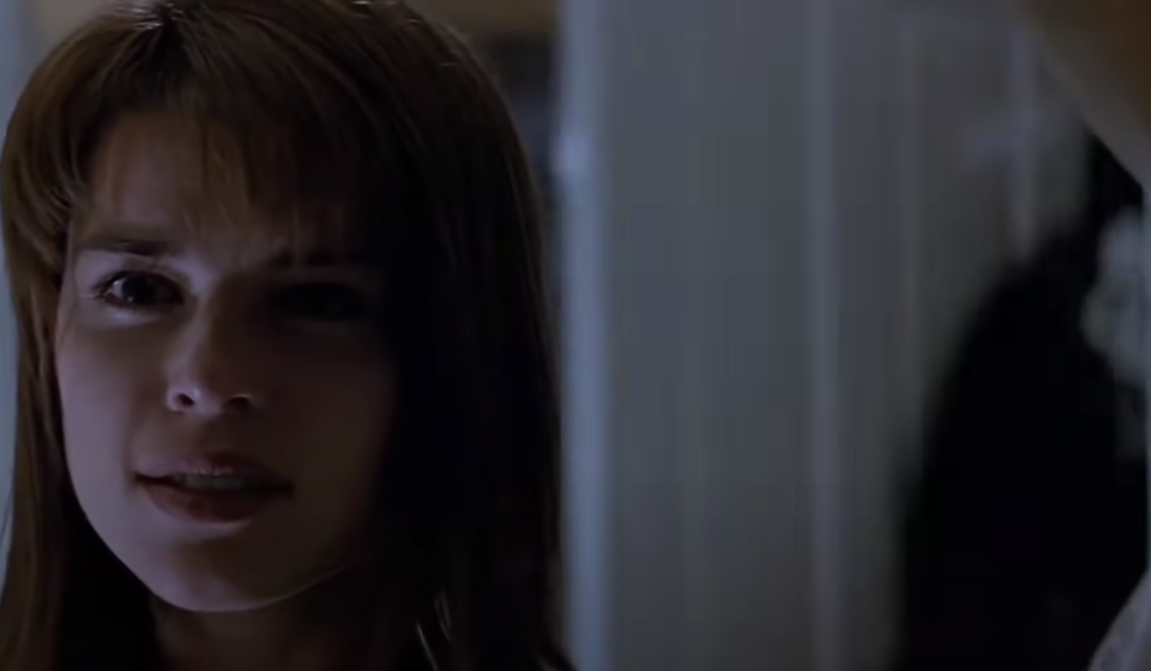} \\
    \includegraphics[height=0.18\textheight]{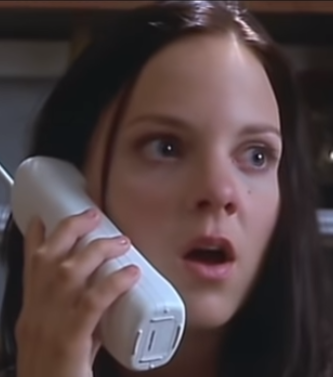}
    \includegraphics[height=0.18\textheight]{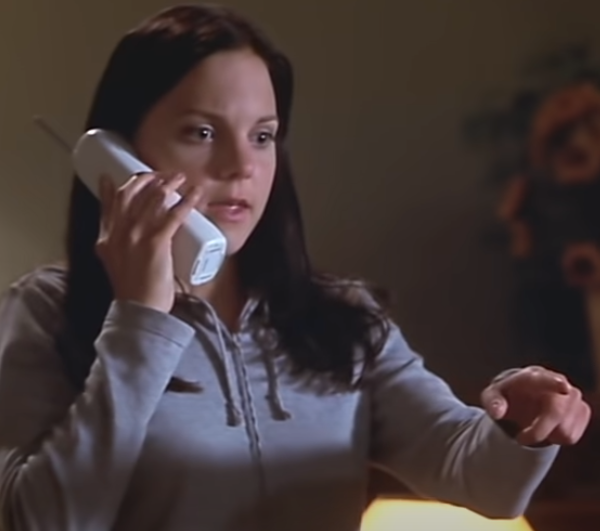} 
    \includegraphics[height=0.18\textheight]{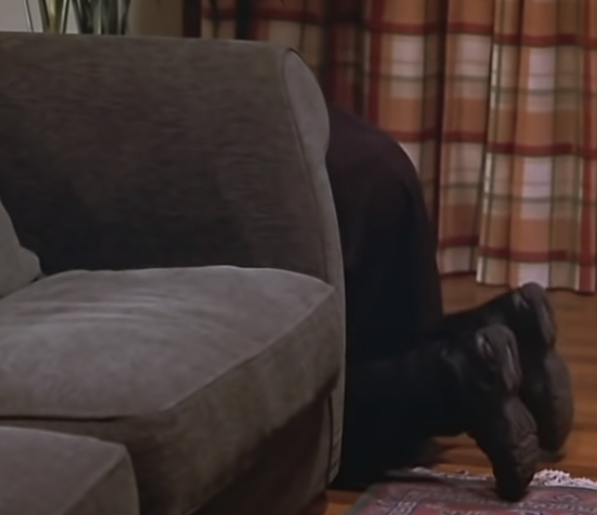} 
    \caption{Some frames from the call scenes of Scream and Scary Movie.}
    \label{fig:scary}
\end{figure}

\begin{itemize}
\item[$K$.] Usually, the caller on the phone is in some remote place.\footnote{This might not be considered true today, but remember that the original script is dated 1995.}
Stalking can escalate to physical violence.
\item[$P$.] A young girl is home alone. She receives a menacing call from a mysterious man who is  likely to be a maniac or a serial killer.
\item[$Q$.] (implicit) Is there a the threat for the girl?
\item[$R$.] (implicit) Yes, somewhere outside.
\item[$R'$.] Yes, hidden inside the house (behind her!).
\item[$\pi$.] Roughly amounts to the difference between two levels of threat, i.e., outside, possibly far away (moderate threat) vs. inside and very close (deadly threat).
\end{itemize}

A generic spectator might even be ``neutral'' toward this scene (e.g., she has neither experienced stalking nor violence).
Still, the overall sentiment is strongly polarized toward fear.
This happens due to the music theme, the fearful expression and voice of Neve Campbell, the narrow close-up, the scarce light, and even the awareness of being watching a horror movie.

Let now consider an alternative version of the protocol.
\begin{itemize}
\item[$K$.] There exists a movie called ``Scream''.
\item[$P$.] A young girl is home alone. She receives a menacing call from a mysterious man who is  likely to be a maniac or a serial killer.
\item[$Q$.] (implicit) Is there a threat for the girl?
\item[$R$.] (implicit) Yes, hidden inside the house (behind her?).
\item[$R'$.] No, an awkward killer hides behind the couch, clearly visible, in front of her.
\item[$\pi$.] Roughly amounts to the difference between two levels of threat, i.e., vicious killer (deadly threat) vs. clumsy killer (low threat).
\end{itemize}

Again, the general sentiment is positively influenced by things such as music, acting style of Anna Faris, relatively wide and bright framing, and (of course) the awareness of being watching a funny movie. 
\end{example}

For what concerns movies and TV shows, directors have many tricks at their disposal to engender the overall sentiment of a scene.
Those mentioned above are just a few examples, but new devices appear time by time.\footnote{For instance, think of the laugh track introduced in the 50s for the first time.}
A similar observation applies to arts in general and to much of our daily experience of the human society.
Since the perception of these devices is limited in a computer, this appears as a major obstacle to the implementation of programs that can recognize emotions at the end of a protocol run.
Nevertheless, it is worth noticing that, in some contexts, the protocol messages carry most of the information about the overall sentiment.
Examples of this fact exist in many jokes and, more recently, in Internet memes~\cite{ling21dissecting}.
Quite often, a joke is a short story carrying some humorous content, which can be told with no specific sentiment device, e.g., it can be written in a book.

\subsection{Types of common knowledge}

Common knowledge is another critical aspect for successfully triggering a certain emotion in the Listener.
As we have seen in the previous section, $K$ may include a lot of complex information that Teller and Listener must own.
As human beings, we have access to a rich cultural background.
Some notions are deeply stratified, e.g., because we gained them during infancy, or strongly rooted, e.g., because we experience them every day.
These categories include our understanding of the rules of the physical world.
Consider the following example.

\begin{example}
In 1949's cartoon ``Fast and Furry-ous'', Wile E. Coyote takes part in the following protocol.

\begin{itemize}
\item[$K$.] Law of impenetrability of bodies.\footnote{Interestingly enough, Chuck Amuck's rule 8 ``Whenever possible, make gravity the Coyote's greatest enemy'' should be rephrased to ``Whenever possible, make laws of physics the Coyote's greatest enemy''.}
\item[$P$.] Wile E. Coyote paints a realistic texture of a tunnel on a solid surface of stone and stays hidden behind a rock, waiting for the road runner.
\item[$Q$.] (implicit) Will the road runner crash against the wall?
\item[$R$.] (implicit) Yes, bodies are impenetrable.
\item[$R'$.] No, the road runner goes through the tunnel.
\item[$\pi$.] Discrepancy between sound physical laws (bodies are always impenetrable) and unsound ones (bodies are sometimes penetrable).
\end{itemize}

The same common knowledge plays a role in a performance by David Copperfield.
In his 1989 special titled ``The Magic of David Copperfield XI: The Explosive Encounter'', the famous magician escapes from a closed safe, inside a building that is about to be demolished.
Eventually, after the building has collapsed, the following protocol takes place.
\begin{itemize}
\item[$K$.] Law of impenetrability of bodies.
\item[$P$.] Copperfield disappeared under tons of rumble. On top of a steel slab, a sheet of orange cloth with a huge, black X lays down, flat.
\item[$Q$.] Where is Copperfield going to appear?
\item[$R$.] Nowhere, he is likely dead.
\item[$R'$.] Right below the X mark, passing through 10 cm of hard steel.
\item[$\pi$.] Discrepancy between sound physical laws (bodies are always impenetrable) and unsound ones (David Copperfield knows how to move across solid objects).
\end{itemize}
\end{example}

Interestingly, protocol failures due to a discrepancy between Teller and Listener knowledge may also occur.
Sometimes, such failures can result in misunderstandings that lead to different emotional states or, noticeably, to the same state but through a different path.
Consider the following example.

\begin{example}
In Futurama season 2 episode 18 ``The Honking'' Bender and the other Planet Express crew members are sitting in the library room of an old, probably haunted mansion.
Blood starts dripping on a wall forming the binary number ``0101100101'', Leela asks Bender ``What does it mean?'' and it answers ``Just gibberish!''.
Then, Bender turns right and sees the number in a mirror. ``1010011010?!?'' it shouts, and flees.

Two different protocol implementations are possible, depending on whether $K$ includes knowledge about binary code.
In case the Listener cannot convert from binary to decimal, the protocol may be the following.
\begin{itemize}
\item[$K$.] Machine (and human) reactions are subject to causal determinism.
\item[$P$.] Bender is annoyed by a gibberish binary sequence.
Bender sees the reverse (still gibberish) binary sequence.
\item[$Q$.] (implicit) How does Bender feel?
\item[$R$.] (implicit) Annoyed.
\item[$R'$.] Terrified.
\item[$\pi$.] Distance between being annoyed and terrified.
\end{itemize}

In case the Listener knows about binary encoding, the protocol may develop as follows.
\begin{itemize}
\item[$K$.] According to human folklore 666 may be scary. Robots are immune to folklore.
\item[$P$.] Bender is annoyed by number $347_2$.
Bender sees $666_2$.
\item[$Q$.] (implicit) How does Bender feel?
\item[$R$.] (implicit) Annoyed, since Bender is immune to folklore.
\item[$R'$.] Terrified.
\item[$\pi$.] Distance between rational robots and irrational human costumes.
\end{itemize}
\end{example}

\subsection{The role of timing}

Another crucial aspect for stimulating high levels of aporia is timing.
The reason is that, as highlighted in most of the previous examples, $Q$ and $R$ are often implicit.
Implicit $Q$ can be conveyed in various ways.
For instance, one can embed it in $P$, e.g., ``How will the coyote fall?'', or by relying on some brain automatism, e.g., ``What is going to happen next?''.

Handling implicit $R$, instead, may be more complex.
Since no synchronization occurs between the Teller and Listener, Teller must estimate the time for the Listener to compute the intended $R$, in a rather accurate way.
On the one hand, too short intervals can prevent the Listener from finding $R$ and ruin the protocol flow.
On the other hand, the Listener might (even partially) anticipate $R'$, so reducing the final aporia level.
Notice that a safety check is carried out in many jokes by starting the protocol with ``Did you hear the one about$\ldots$?''.

\begin{example}
In 1963's cartoon ``To Beep or not to Beep'', Wile E. Coyote resorts to a catapult.
Interestingly, this results in a sequence of gags, all following the protocol below.

\begin{itemize}
\item[$K$.] Machines are subject to causal determinism.
\item[$P$.] Wile E. Coyote pulls the rope that triggers the catapult. Each time he changes his position (for safety).
\item[$Q$.] (implicit) How will the catapult hurt the coyote?
\item[$R$.] (implicit) As before, machines are deterministic.
\item[$R'$.] Each time in a different way.
\item[$\pi$.] Discrepancy between previously observed behaviours and the current one.
\end{itemize}

\begin{table}[t]
    \centering
    \begin{tabular}{r r r r r}
        Safe position & Scene start & Rope pulled & Crushing & Scene end  \\
        \hline
        Behind (close) & 0.00 s & 3.75 s & 4.83 s & 6.61 s \\
        Front & 8.20 s & 11.09 s & 11.72 s & 14.11 s \\
        Behind (far) & 15.16 s & 18.59 s & 19.27 s & 21.62 s \\
        Right & 22.35 s & 23.80 s & 24.73 s & 26.61 s \\
        Under & 27.68 s & 30.56 s & 31.28 s & 34.59 s \\
        \hline
        Average step & 0.00 s & +2.88 s & +0.81 s & +2.34 s \\
        \hline
    \end{tabular}
    \caption{Catapult gags duration (source \url{https://youtu.be/bmEGpNCYuRY}).}
    \label{tab:catapult}
\end{table}

Reasonably, each time the protocol is repeated, i.e., for each gag, the choice of timing is subject to the same rules.
Table~\ref{tab:catapult} reports the time of the key events of each gag.
The time interval between the scene start (black screen fades out) and the crushing event is used by the Listener to compute $R$.
By analysing the scene before the rope is pulled, one may anticipate $R'$ if enough time is given.
The longest intervals before rope pulling are for gags 1 ($3.75 s$) and 3 ($3.43 s$).
Since at the first gag the spectator has no experience of what might go wrong with the catapult, anticipating the outcome is more difficult.
Interestingly, in the third gag, Will E. Coyote is squashed by the catapult itself (rather than the stone), which might be unexpected and, thus, require more time for elaboration. 
Finally, it is worth noticing that, after the rope is pulled, the crushing event follows quickly ($0.81 s$ on average).
This may be due to the fact that the dynamics of object, e.g., the stone trajectory, reveals much information about $R'$.
Remarkably, $0.81 s$ is very close to the \emph{substantial pause}, i.e., $[0.6 s - 0.8 s]$, defined in~\cite{Brown80questions} and used in~\cite{Attardo11timing} for signaling an upcoming punch line in verbal jokes. 
\end{example}

Time can even have an active role in the narration of the protocol, i.e., time can appear as part of $P$ or $Q$.
In these cases, the Listener can take it into account when computing $R$.
Curiously, there are even cases in which longer pauses play a role.
This happens when a short pause might make the Listener opt for $R'$ (or an answer that is close), while a long pause may lead to the $R$ that Teller is aiming at.
Consider the following, classical example.

\begin{figure}[t]
    \centering
    \includegraphics[height=0.16\textheight]{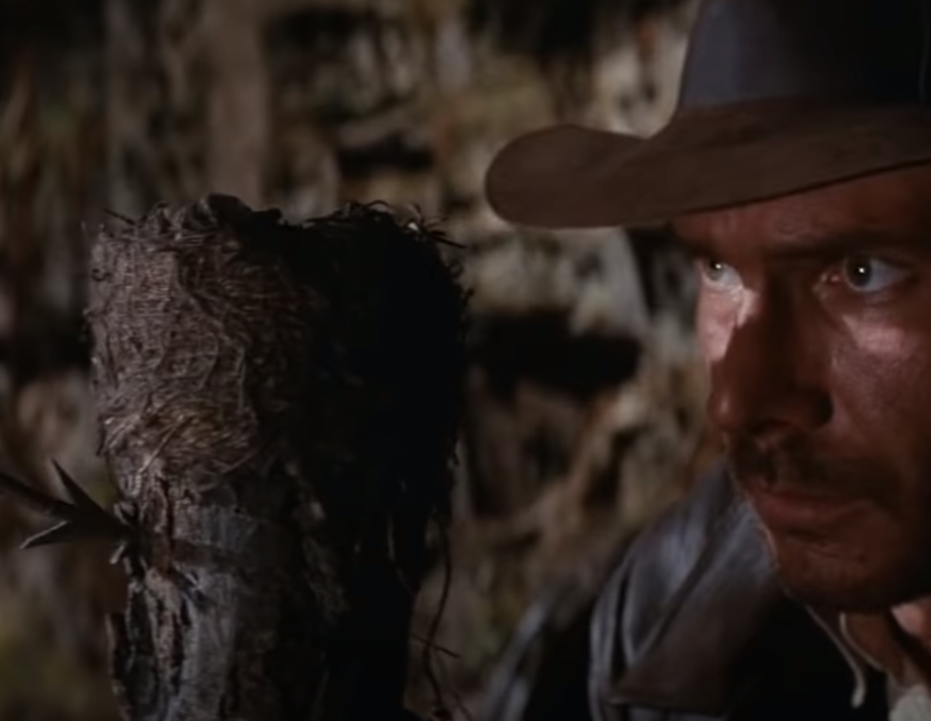}
    \includegraphics[height=0.16\textheight]{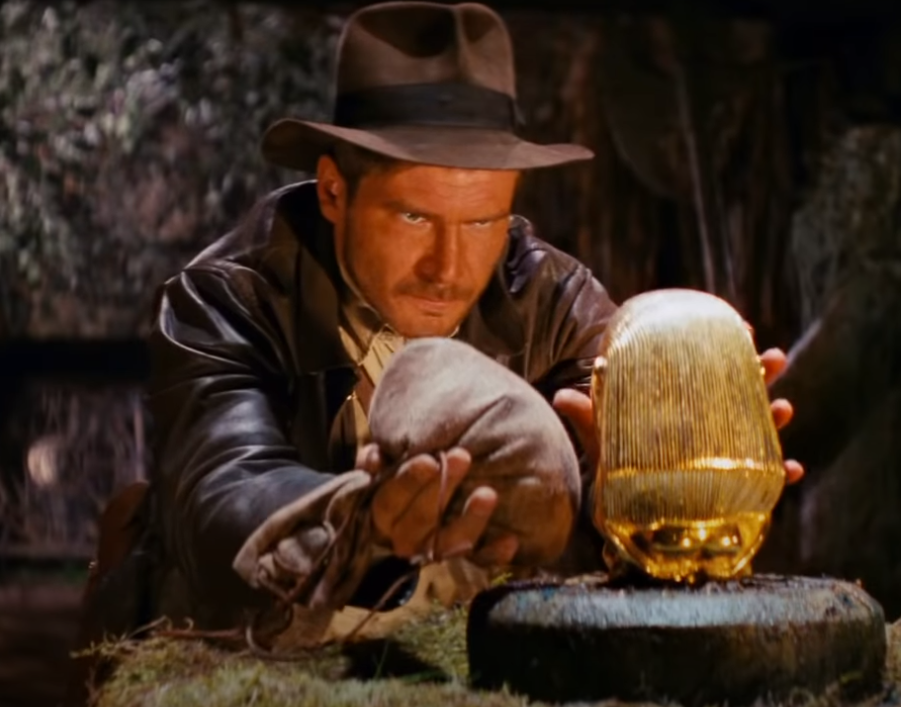}
    \includegraphics[height=0.16\textheight]{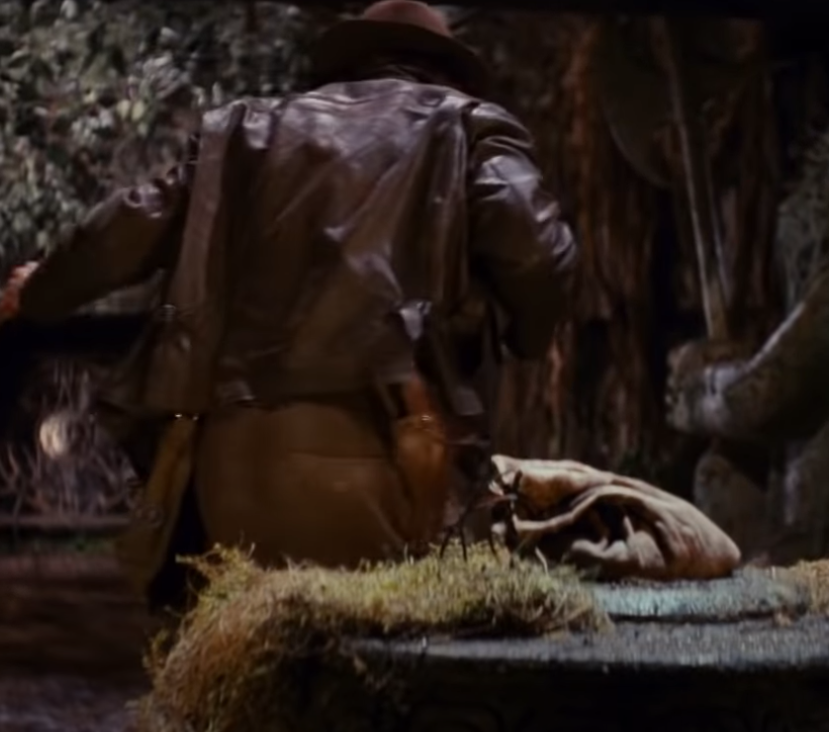}
    \caption{Indiana Jones triggers the pedestal switch of the Chachapoyan idol.}
    \label{fig:indy}
\end{figure}

\begin{example}
\label{ex:indy}
In 1981's ``Indiana Jones and the Raiders of the Lost Ark'', Indiana Jones (Harrison Ford) and his guide Satipo (a young Alfred Molina) are trying to steal the fertility idol from the temple of the Chachapoyan Warriors.
When they eventually get to the idol chamber, an extremely tensive scene takes place (see Figure~\ref{fig:indy}).
We can represent the scene through the following protocol.

\begin{itemize}
\item[$K$.] The temple has quick\footnote{E.g., arrows are fired immediately when a plate is pressed.} deadly traps.
Indiana Jones can avoid traps he is aware of. 
Indiana Jones is aware of the pedestal trap.\footnote{Indeed, he filled a sack of sand right before entering the temple.}
\item[$P$.] Indiana carefully estimates the weight of the idol and removes some sand from the sack.
He cautiously replaces the idol with the sack.\footnote{In the meanwhile Satipo nervously rubbing his fingertips remarks the growing suspense.}
Then, Indiana holds the breath watching at the pedestal for an instant before smiling\footnote{Also Satipo smiles.} and turning back.
\item[$Q$.] (implicit) Did Indiana deactivate the trap?
\item[$R$.] (implicit) Yes, since traps are quick and a few seconds passed.
\item[$R'$.] No, the pedestal slowly starts to move downward.
\item[$\pi$.] Indiana is safe vs. Indiana is in danger.
\end{itemize}
\end{example}

Before moving to the next topic it is important to remark that timing historically represented a difficult matter in the study of humor.
This is stressed in~\cite{Attardo11timing}, where the following three definitions of timing are given.

\begin{enumerate}
    \item Timing as distribution of pauses. 
    \item Timing as distribution of text (aka rhythm).
    \item Timing as interaction with others.
\end{enumerate}

All of these aspects are naturally present in protocols where two parties (3) exchange messages consisting of arbitrary text (2) by interleaving them with arbitrary pauses (1).

\subsection{Computing distances}

Understanding and estimating the distance between $R$ and $R'$ is fundamental.
As a matter of fact, distance is the main factor that drives the intensity of the emotion perceived by the Listener.
Defining distances is non trivial in general.
The reason is that aporia often occur over non-metric domains, e.g., in most of the previous examples aporia arises from logical inconsistencies.
Nevertheless, the notion of distance is more general than that of inconsistency.
On the one hand, inconsistency can be modeled as the distance between two far or extreme positions.
On the other hand, some jokes rely on distant, yet coherent, propositions.
For instance, this happens in exaggeration jokes in which the Listener can anticipate the general structure of the answer, but she cannot figure out its intensity.

\begin{example}
Consider the following fragment taken from ``Hilarious'' (2010) by Louis C. K.
\begin{quote}
We don’t think about how we talk. 
We just say, `Dude, it was amazing. It was amazing.' Really? You were `amazed'? You were `amazed'? By a basket of chicken wings? Really? `Amazing'?

What are you going to do with the rest of your life now? 
What if something really happens to you? 
What if Jesus comes down from the sky. 
Makes love to you all night long. 
Leaves the new living lord in your belly?    
\end{quote}

In both the previous jokes, the Listener can anticipate that Louis C. K. is going to make an example of something barely amazing and very amazing, respectively.
However, the distance between $R$ and $R'$ is remarkable.
Indeed, in the first example $R'$ is just ordinary (``a basket of chicken wings'').
Even more, in the second example, $R'$ consists of an exaggeration built through a few steps that progressively exacerbate the overall level of ``amazingness''.
\begin{enumerate}
    \item Jesus comes down from the sky
    \item Makes loves to you all night long
    \item Leaves the new living lord in your belly
\end{enumerate}
\end{example}

A further, relevant fact is that, in many contexts, human beings are not very good at accurately estimating values.
Also, often subjectivity comes into play.
Consider the following example.

\begin{example}
\label{ex:balance}
Ron checks his balance online before transferring $n$\$ to Elliot.
The balance is $N$\$ (with $N > n$).
Now imagine that Ron runs this protocol, where the Teller is his e-Banking service.

\begin{itemize}
\item[$K$.] Balance is $N$\$.
\item[$P$.] Ron transfers $n$\$. Ron checks his balance again.
\item[$Q$.] (Implicit) What is the new balance?
\item[$R$.] (implicit) $(N - n)$\$.
\item[$R'$.] The new balance is $0$\$.
\item[$\pi$.] Distance between $(N-n)$ and $0 = \Vert N - n - 0 \Vert = N - n$.
\end{itemize}

Although $\pi = N - n$ may seem reasonable, it is likely that the final level of aporia is related to the actual values of $N$ and $n$, as well as the relative impact of the loss on Ron's wealth (see below).
\end{example}

Instead of metric spaces with measurable distances, in many cases humor (as well as other emotions) is triggered by a conflict between two \emph{theories}, i.e., a system of axioms or a personal belief, under which the same fact can be interpreted.  
In particular, aporia may arise when there are strong differences between the two theories.
For instance, a certain fact may be proved true and false, respectively, or there is a strong difference between the proofs developed under each theory.
Also, as already mentioned, a certain belief can be deeply rooted in the Listener and a significant effort is necessary in order to abandon such a theory, e.g., think of common sense or dogmatism.
If the Listener refuses to give up with one of the two theories, the aporia effect may be nullified.
Comedians that treat sensitive subjects often face similar problems.

\begin{example}
\label{ex:hicks}
The brilliant stand-up comedian Bill Hicks, during his show ``Relentless'', tells a joke\footnote{Available at \url{https://vimeo.com/490599763}.} corresponding to the following protocol.

\begin{itemize}
\item[$K$.] People may get offended when their belief is mocked (1). Christian belief must imply forgiveness (2).
\item[$P$.] After a show, Bill is faced by three rednecks (sic) who threaten him. They say to be christian and to be offended by Bill's jokes.
\item[$Q$.] (implicit) Are they right?
\item[$R$.] (implicit) Yes, by (1).
\item[$R'$.] No, by (2).
\item[$\pi$.] $R$ implies $\neg$ (2), i.e., abandoning a strong belief, while $R'$ doesn't.
\end{itemize}

Notice that, in this example, logic modalities \emph{may} and \emph{must} play a crucial role. 
As a matter of fact, $R'$ also complies with (1), i.e., getting offended is not mandatory.
\end{example}

\begin{figure}
    \centering
    \includegraphics[height=0.18\textheight]{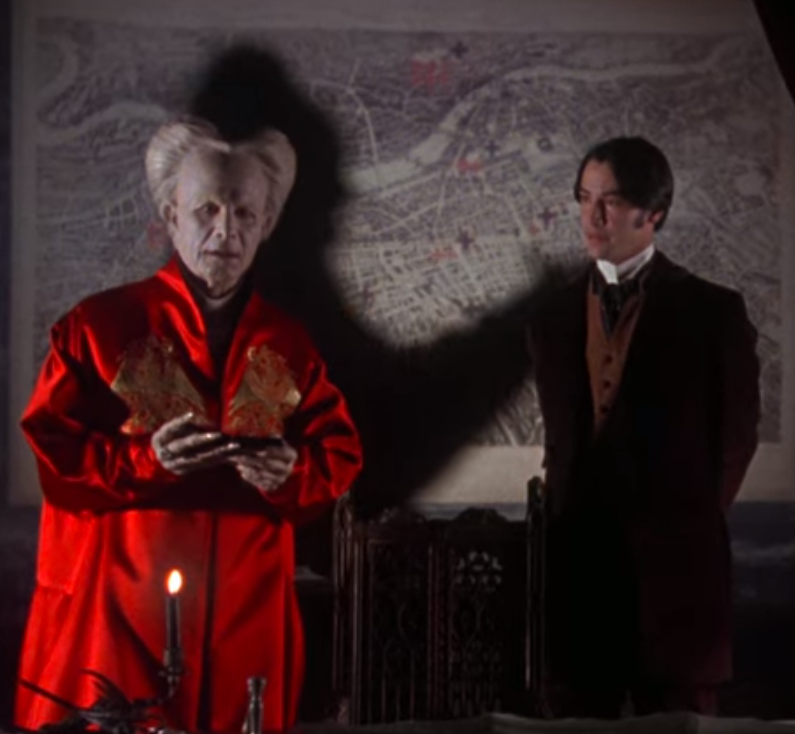}
    \includegraphics[height=0.18\textheight]{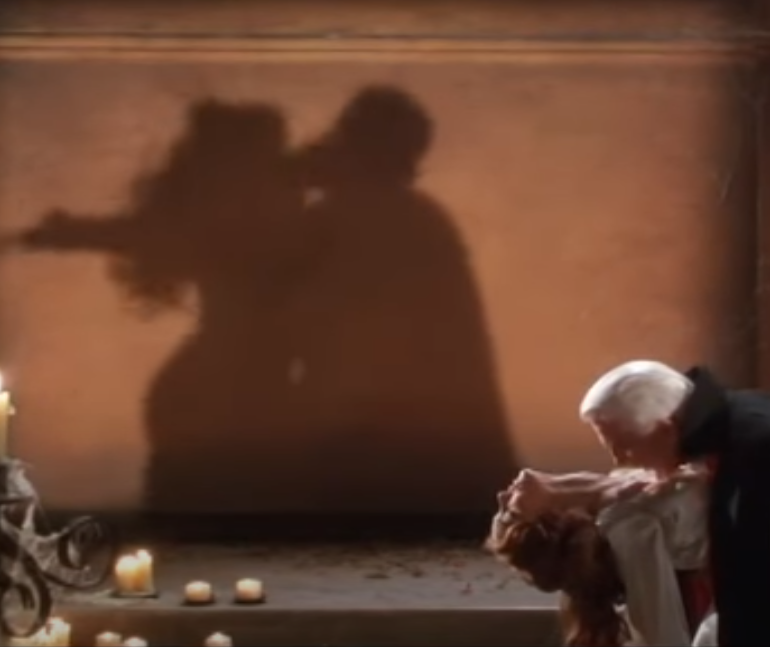}
    \includegraphics[height=0.18\textheight]{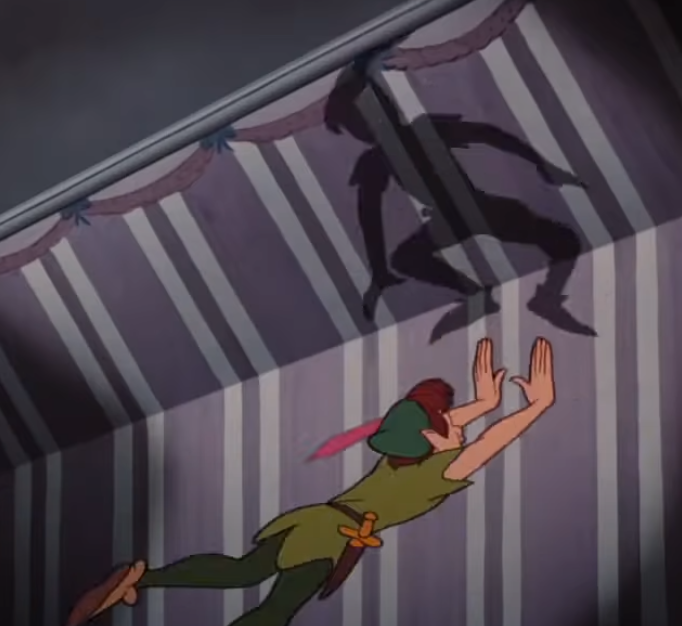}
    \caption{Anomalous shadows in movies.}
    \label{fig:shadow}
\end{figure}

Violations of the laws of physics and common sense are also source of high levels of aporia.
We already mentioned Will E. Coyote and his bad luck with classical mechanics.
Light and shadows are also familiar in everyone's experience and a violation of the expected behavior may result in a strong sense of aporia, e.g., think of ``Bram Stoker's Dracula'', ``Dracula: Dead and Loving it'' or ``Peter Pan'' (see Figure~\ref{fig:shadow}).

Logic rules and inferences are also very effective.
For instance, redundancy and superfluous assumptions are good to stress the distance between two theories.

\begin{example}
Jack Benny's famous joke ``Give me golf clubs, fresh air and a beautiful partner, and you can keep the clubs and the fresh air'' can be modeled as follows.

\begin{itemize}
\item[$K$.] Golf clubs, fresh air and a partner are needed for playing golf. A beautiful partner is needed for more pleasant activities.
\item[$P$.] Jack Benny asks for golf clubs, fresh air and a beautiful partner.
\item[$Q$.] (implicit) What for?
\item[$R$.] (implicit) Playing golf.
\item[$R'$.] Returning golf clubs and fresh air (and dedicating to more pleasant activities).
\item[$\pi$.] $R'$ is redundant (w.r.t. premises in $P$) while $R$ is not.
\end{itemize}
\end{example}

Other theories such as social conventions, traditions, superstitions, education, etc. can be added to these examples.
Intuitively, in these cases aporia seem to be related to a \emph{cost function} $\gamma$ which the Listener computes to estimate the value of involved theories.
As a matter of fact, theories represent a fundamental asset for an intelligent being.
Rejecting a theory that we formed after many experiments and that seems to be working well, e.g., ``things fall from top to bottom'', may shake the way we perceive and understand phenomena.
Thus, it should not be done lightly.
In general, we can also think that an individual shall not even consider rejecting a theory $\varphi$, e.g., a religious dogma, when its cost surpasses a certain threshold $T$, in symbols $\gamma(\varphi) > T$.\footnote{For the sake of presentation, from now on we assume $\gamma(\varphi)$ and $T$ to range in $[0,1]$.}
In these cases, the aporia level falls to 0.
The resulting aporia protocol is as follows.

\begin{itemize}
\item[$K$.] Listener believes in $\varphi$.
\item[$P$.] A statement that can be interpreted under both $\varphi$ and $\neg \varphi$.
\item[$Q$.] A question that is true under $\varphi$ and false under $\neg \varphi$.
\item[$R$.] true.
\item[$R'$.] false.
\item[$\pi$.] $ = \left\{\begin{array}{l l}\gamma(\varphi) & \textnormal{ if } \gamma(\varphi) < T \\
0 & \textnormal{ otherwise}\end{array}\right.$
\end{itemize}

It is not difficult to check that the joke by Bill Hicks, given in Example~\ref{ex:hicks}, follows this scheme.
Subjective choices of $\gamma$ and $T$ model the intuitive notions of ``skepticism'' and ``open mindedness''.
Thus, a skeptical (high $T$) and open minded (low $\gamma$) person may consider the joke witty, while a dogmatic (low $T$), narrow minded (high $\gamma$) individual would find it unfunny.
To better highlight this behavior, consider the following variant of Example~\ref{ex:balance}.

\begin{itemize}
\item[$K$.] Ron is a wealthy man (e.g., $N = 100000$).
\item[$P$.] Ron transfers $100$\$. Ron checks his balance again.
\item[$Q$.] (implicit) What is the new balance?
\item[$R$.] (implicit) $99900$\$.
\item[$R'$.] The new balance is $0$\$.
\item[$\pi$.] $\gamma($Ron is less wealthy than before$)$ (or $0$ if $T$ is surpassed).
\end{itemize}

Clearly, Ron might tend to reject the new theory, while an external observer (having a different $\gamma$) can find it less difficult to switch.
On the opposite, if after withdrawing Ron discovers that the balance has doubled, he might be open to accept the new theory, i.e., ``Ron is wealthier!''.
In general, it appears reasonable that, when two or more alternative theories can explain $R'$, the Listener subjectively opts for the less costly.
Consider the following case.

\begin{example}
The assumption ``intelligent beings will act rationally'' appears very reasonable and most people might believe it to be true in general.
Roughly, this means that intelligent/rational beings will always opt for the highest gain (or lowest loss) when taking strategic decisions.
Now consider the following statement ``morphology and physiology of anthropomorphic beings resemble human ones''.\footnote{E.g., if something looking like a nose is in the middle of something looking like a face, it is likely to be used for breathing and smelling things.} Even though it looks more arguable, we might be slightly prone to accept it, at least for symmetry and since we have seen many examples in nature and in sci-fi movies.
In Rick and Morty season 3 episode 8 ``Morty's Mind Blowers'', Rick and Morty interrogate an alien to obtain secret codes (unlocking vaccines, otherwise ``Earth dies''!).
The alien speaks an incomprehensible language that Rick translates.
The scene is depicted in Figure~\ref{fig:rick}, and it follows the protocol below.

\begin{figure}
    \centering
    \includegraphics[height=0.16\textheight]{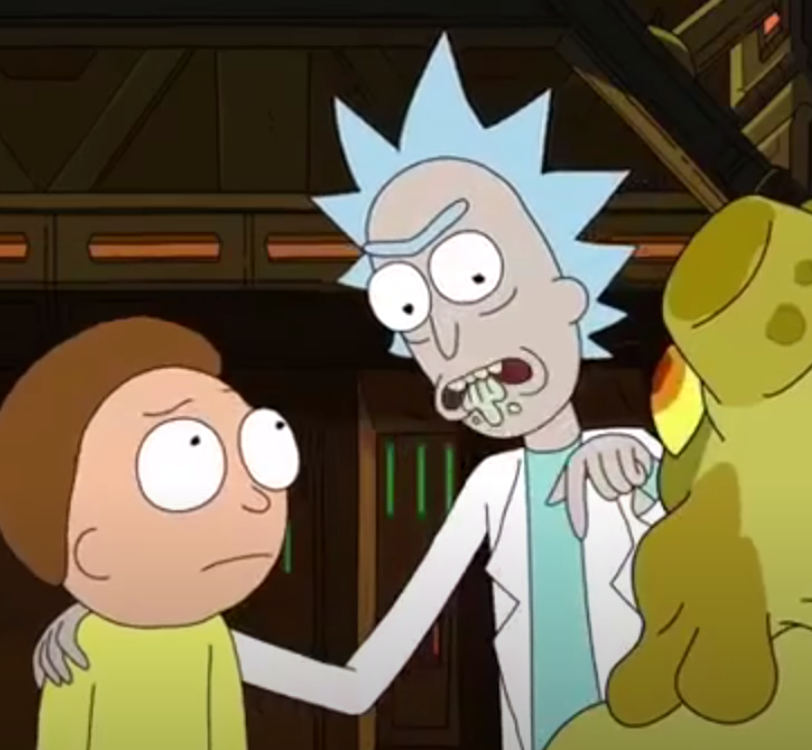}
    \includegraphics[height=0.16\textheight]{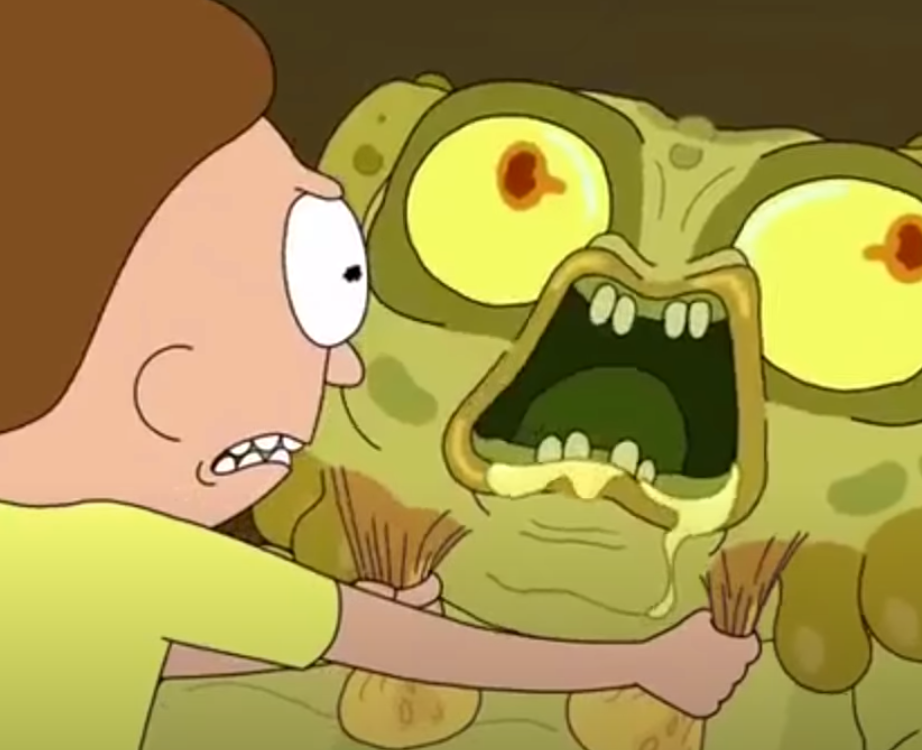}
    \includegraphics[height=0.16\textheight]{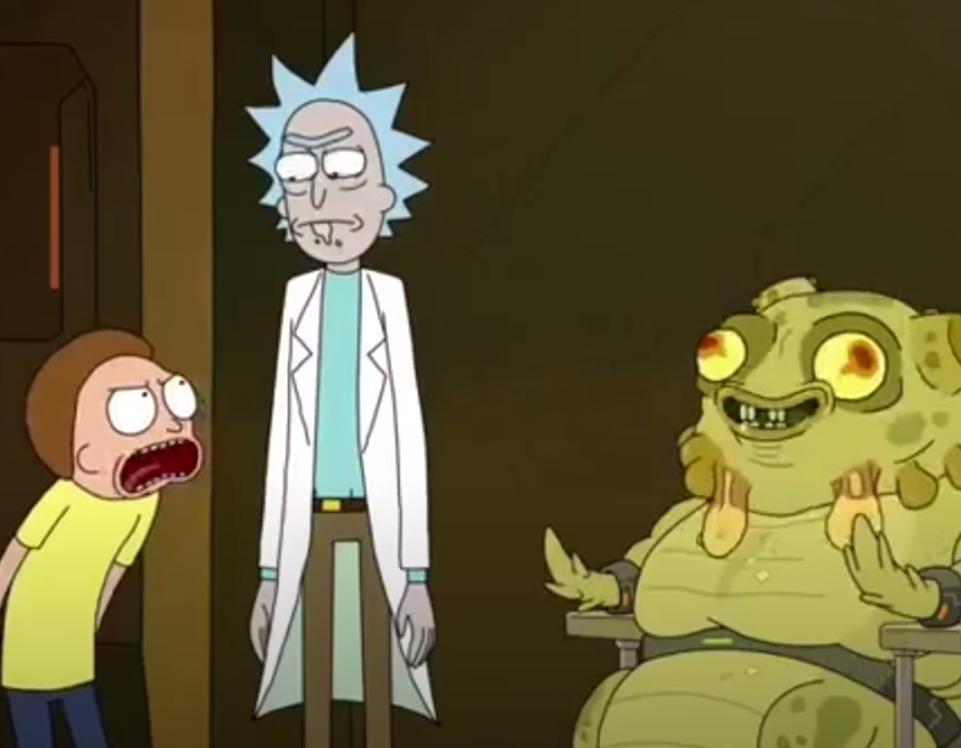}
    \caption{Rick and Morty interrogating the alien.}
    \label{fig:rick}
\end{figure}

\begin{itemize}
\item[$K$.] Intelligent beings will act rationally (1). Morphology and physiology of anthropomorphic beings resemble human ones (2).
\item[$P$.] Rick and Morty interrogate the alien who refuses to collaborate.
Visibly angry, Morty insists for ``step up'' the interrogation.
Rick tells Morty to grab and twist the ``flashy sacs under his chin''.
The alien screams, Rick says to stop the treatment and the alien proposes to make a deal. 
\item[$Q$.] (implicit) What is the deal?
\item[$R$.] (implicit) The codes for stopping the treatment.
\item[$R'$.] The codes for continuing the treatment (``Half the codes now, half after you finish'').
\item[$\pi$.] $\gamma($2$)$ (i.e., Morty is sexually stimulating the alien, instead of torturing it).
\end{itemize}

In this case, $R$ is justified by (1) $\land$ (2), i.e., the alien makes a rational choice to stop the pain.
Instead, $R'$ can be justified in two ways.
The straightforward one (given in the protocol) is for (1) $\land$ $[\neg$(2)$]$, i.e., the alien makes a rational choice to continue the pleasure.
An alternative one would be for $[\neg$(1)$]$ $\land$ (2), i.e., the alien makes an irrational choice to continue the pain.
However, the second interpretation would require invalidating (1) instead of (2).\footnote{Notice that other signals, such as the smiling expression of the alien, only occur after $R'$ has been delivered. As a matter of fact, when Rick translates the deal, the framing goes to a close up of Morty's face. See \url{https://youtu.be/OtsA8j3Tv1E}.} 
The resulting aporia level derives from the distance between Morty's original belief and reality.\footnote{Also, it is exacerbated by the sexual allusion.}
\end{example} 

\subsection{Aporia in static artwork}

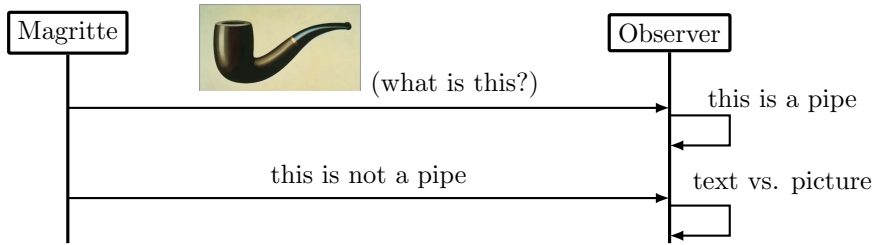
\begin{figure}[t]
  \centering
  
  \begin{tikzpicture}[every node/.append style={very thick,rounded corners=0.2mm}]
    
    \node[draw,rectangle] (Prover) at (0,0) {Magritte};
    
    \node[draw,rectangle] (Verifier) at (8,0) {Observer};
    
    \draw [very thick] (Verifier)--++(0,-2.8);
    \draw [very thick] (Prover)--++(0,-2.8);
    
    \draw [-latex,thick] (0,-1)--node [auto] {\includegraphics[height=0.06\textheight]{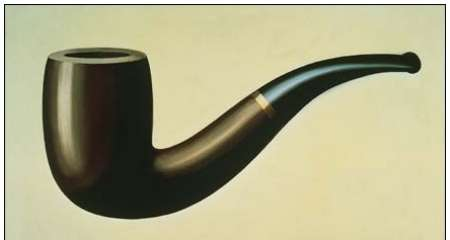}~(what is this?)}++(8,0);
    
    
    \draw [-latex,thick] (0,-2.2)--node [auto] {this is not a pipe}++(8,0);
    
    \node at (9.5,-0.9) {this is a pipe};
    \draw[-latex,thick] (8,-1.1) -- +(0.8,0) |- (8,-1.5);
    
    
    \node at (9.5,-2) {text vs. picture};
    \draw[-latex,thick] (8,-2.3) -- +(0.8,0) |- (8,-2.7);
    
    
    \end{tikzpicture}

    \caption{Aporia protocol for ``Ceci n'est pas une pipe''.}
    \label{fig:pipe}
\end{figure}

As already said, many artists showed interest for incongruity as a trigger for various emotional states.
In the previous examples we always considered dynamic artwork, e.g., movies, cartoons, books and stand-up comedy.
These examples embed an explicit notion of time, through which the protocol flows, thus making the protocol simpler to be identified and modeled.
Needless to say, designing an aporia protocol without an explicit notion of time in more challenging.
A possible solution is to rely on perception automatism that are shared by most individuals.
For instance, many people tend to process pictograms before text.
Also, in most cultures, visual inputs are processed from top to bottom.
A prominent example of this automatism is provided by ``Ceci n'est pas une pipe'', by the Belgian surrealist Ren\'e Magritte.
The protocol between the painting and the observer may be modeled as in Figure~\ref{fig:pipe}.

\begin{figure}[t]
    \centering
    \includegraphics[height=0.22\textheight]{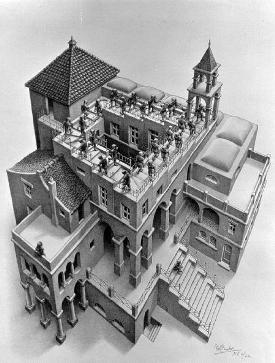}
    \includegraphics[height=0.22\textheight]{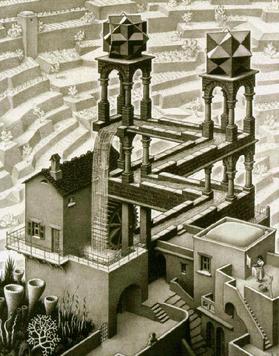}
    \includegraphics[height=0.22\textheight]{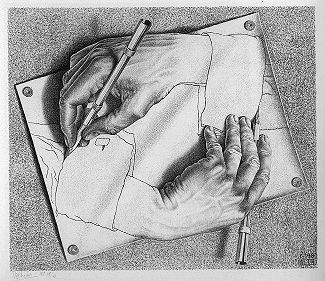}
    \caption{``Ascending and Descending'', ``Waterfall'' and ``Drawing hands'' by M. C. Escher.}
    \label{fig:escher}
\end{figure}

Other devices include 3D perspective and recursion, as in M. C. Escher's artwork. 
His lithographs ``Ascending and Descending'', ``Waterfall'' and ``Drawing hands (see Figure~\ref{fig:escher}) well represent this effect.
All these pictures are meticulously detailed, and an observer may need a few instants to process them entirely.
Reasonably, different parts will be analyzed in a sequence.
Interestingly, such a sequence can be influenced by the picture itself.
For instance, when a stream of flowing water is identified, we might be tempted to follow it and, thus, run an aporia protocol that concludes with a paradox (the stream forms a loop).
To increase the overall level of aporia, interestingly, most other elements in the pictures respect the prospective conventions.
In this way, the observer can get to $R'$ without expecting it. 

\section{Testing emotions}
\label{sec:poe}

In the previous section we have discussed the abstract structure of aporia protocols and their relationship with human emotions.
The reason is that most of our experience is about human interactions.
Nevertheless, many open research problems are strongly related to our methods for testing emotional transfers.
In this section we present some applications of the aporia protocol model that refer to research problems of interest.

\subsection{Proof of Emotion Test}

We propose a new version of the classical TT, called \emph{Proof of Emotion Test} (PoET).
As for TT, PoET consists of an interview, where a human interviewer interacts with an agent.
The tested agent can be either another human being or a computer program.
The interviewer and the tested agent only communicate through a channel that cannot reveal their identity, e.g., using a computer terminal.
The main novelty is that the interviewed agent claims to understand human emotions and the interviewer has to verify it.
The test follows these steps.

\begin{enumerate}
\item The interviewer and the agent agree on a finite set of emotions $E = \{ e_1, e_2, \ldots, e_k \}$.
\item The interviewer takes the role of the Teller in the aporia protocol, whereas the agent plays the role of the Listener.
\item The interviewer sends a pair $(P,Q)$, where $Q$ is an explicit question.
\item The agent sends back an answer $R$ and the interviewer replies  with $R'$.
\item the agent communicates which element $e_i \in E$ better describes the emotion stimulated by the last aporia protocol.
\item If the interviewer is ready to emit the final verdict, the test is interrupted. Otherwise, another session of the protocol is executed (step 2).
\end{enumerate}

\subsection{An algorithm for playing PoET}

\begin{figure}[t]
    \centering
    \includegraphics[width=0.98\textwidth]{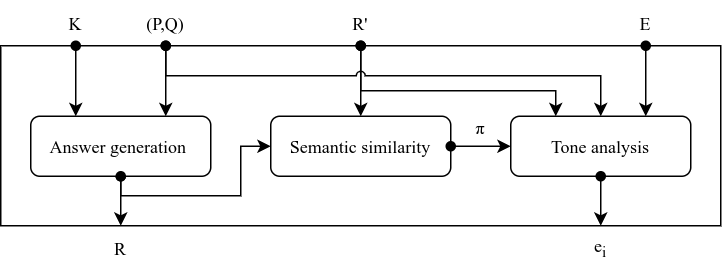}
    \caption{Architecture of a PoET algorithm.}
    \label{fig:architecture}
\end{figure}

In Figure~\ref{fig:architecture} we sketch the structure of an algorithm for playing PoET against a human interviewer.
From left to right, the algorithm starts by processing inputs $(P,Q)$, possibly under knowledge $K$, and generates the answer $R$.
When the interviewer submits $R'$, it is compared with $R$ to check semantic similarity.
The resulting aporia level $\pi$ can be obtained, for instance, as the inverse of the similarity score.
Eventually, $(P,Q)$, $K$, and $R'$ are analyzed to identify the prevailing tone.
By composing the sentiment and the aporia level $\pi$, an emotion $e_i$ is identified from set $E$. 

Interestingly, the building blocks needed to implement a program for playing PoET may already exist.
For instance, IBM Watson already include most of them.\footnote{See \url{https://cloud.ibm.com/developer/watson/documentation}.}

\subsection{Searching for emotion-based interactions}

It is well known that most animals feel emotions.
The reason is that many emotional states are associated with observable, sometimes measurable, physiological and behavioral signals.
An animal typically uses these signals to alert others about its own emotional state and, sometimes, to stimulate others' emotions.
For instance, a dog showing teeth as a sign of anger probably wants to trigger fear in its opponent.
Recognizing emotional signals is somehow instinctive and even works between different species, e.g., think of a pray taking defensive stance.

Aporia protocols, instead, can be used to communicate a specific emotional state in a fully symbolic way.
In this respect, observing aporia protocols in action between animals might be a sign that complex emotional information is transferred by the Teller and, possibly, understood by the Listener.
This is a necessary condition, e.g., for empathetic behavior.

\subsection{Introducing emotions in computation}

A less studied, yet interesting question is whether some sort of emotions can exist beyond natural ones.
Said differently, can we implement computer interactions, based on aporia protocols, regardless of whether they trigger human emotions or not?
Also, and perhaps more importantly, do they have any application of interest?

In some sense, trust management systems~\cite{Weeks01tms} already encode a sort of ``emotional'' state, e.g., in service oriented computing (SOC)~\cite{Papazoglou03soc}.
SOC environments consist of several services that can take part in complex, distributed applications.
If two or more alternatives exist for completing a task, services are ranked according to their features, namely their \emph{contract}~\cite{Leitner07contract}.
Contracts can be deduced from previous interactions or exposed by the service itself.
In any case, they are not perfectly reliable, e.g., since deduction can be approximated or the service may be cheating.

Often the overall level of trust associated with a service is reduced when some computation is observed that violates the expected contract.
For instance, consider again the scenario of Example~\ref{ex:balance}.
Now imagine that Ron is a service invoking a function \verb|transfer| $: \mathbb{N} \rightarrow \mathbb{N} \times \mathbb{N}$, provided by the e-Banking service.
Intuitively, \verb|ob, nb := transfer(a)| transfers \verb|a|\$\footnote{The recipient is immaterial for this example.} and returns both the old balance \verb|ob| and the new one \verb|nb|.
The contract that Ron associates to this function says that 
\[
\mathtt{nb} = \left\{\begin{array}{l l}
     \mathtt{ob - a} & \textnormal{ if } \mathtt{ob \geq a} \\
     \mathtt{ob} & \textnormal{ otherwise }
\end{array}\right.
\]
In terms of reliability of this contract, both \verb|100, 40 := transfer(50)| and \verb|100, 60 := transfer(50)| may be considered as violations.
However, different emotions might by triggered by these events.
In general, as for trust-based policies, one could even define rules that take into account a complex emotional state.
Different emotions can be associated with different types of resources, e.g., computation time, money, private data, etc.
Whenever a contract is violated, a specific emotion is triggered depending on the level of aporia and the type of involved resources.
Then, emotion-based policies are used to drive the next interactions.
For instance, a policy may state that invocations should only be done toward services we are \emph{happy} with (e.g., actual cost was less or equal than expected one) and we are not \emph{bored} of (e.g., actual computation time was at most 120\% of expected one).

\section{Related work}
\label{sec:related}

In different disciplines, many authors proposed models for describing how emotions are stimulated during input processing.
Nevertheless, to the best of our knowledge, our proposal is the first one considering two-party protocols in which a Teller intentionally transfers emotions to a Listener.
Nevertheless, some of the key elements of our proposal are strongly related to previous works.

\emph{Incongruity theory}, mainly applied to humor, has a longstanding tradition.
It was originally proposed by Scottish poet James Beattie (1779) and it inspired the work of many philosophers, including Immanuel Kant, Arthur Schopenhauer and S\o ren Kierkegaard.
The overall idea that humor is triggered when incongruity is observed and processed by an observer had a considerable impact in several fields, e.g., including marketing~\cite{Alden00effects}.
Furthermore, recently incongruity theory has received biological support by studies on the role of amygdala in cognitive processes~\cite{Nakamura18amygdala}.
We remark that incongruity theory is not in contrast with our proposal.
In fact, our notion of aporia as a measure of the distance between two propositions, can be interpreted as an operational definition of incongruity. 
\emph{Relief theory} and \emph{superiority theory} are also of great interest. 
Briefly, the former states that humor arises when some stress is eventually and suddenly released, while the latter states that laugh originates from a sense of superiority w.r.t. someone else.
As for incongruity theory, these proposals are not in contrast with our protocol model.
In particular, they can be seen as types of distance functions (``how much do I feel relieved/superior''?) that are common for human beings.

Although not rigorously defined, some authors considered the role of \emph{trajectories} for the development of jokes.
For instance, this fact was emphasized by John A. Paulos~\cite{Paulos82math}, who proposed catastrophe theory to model humor.
His main example is taken from~\cite{Zeeman73catastrophe} where emotional indicators, e.g., the position of ears, are measured in dogs.
In his proposal, trajectories amount to paths leading to the catastrophe, i.e., a cusp of the surface describing the joke.
Nevertheless, no precise definition of these trajectories is given in mathematical terms.

A more computational approach is followed by~\cite{suslov2007computer}, where trajectories consist of alternative continuations of a sequence of symbols, based on a probability distribution.
This proposal introduces a notion of temporal development of the joke as the flow of input symbols processed by a reader.
In this model, the humorous effect is said to arise from the difference between anticipations and actual symbols.
Such a difference occurs due to the reader looking ahead for symbols that are more likely to follow last observed ones.
Again, no clear definition of trajectory is provided.
Also, the proposal of~\cite{suslov2007computer} has no notion of distance or incongruity between propositions.

The General Theory of Verbal Humor (GTVH), a very influential linguistic theory, was proposed by Attardo and Raskin in~\cite{Attardo91humor}.
Briefly, GTVH treats jokes in terms of six knowledge resources, namely \emph{script opposition}, \emph{logical mechanism}, \emph{situation}, \emph{target}, \emph{narrative strategy} and \emph{language}.
Their proposal can be combined with the model presented in this work, e.g., for the linguistic characterization of $P$, $Q$, $R$ and $R'$.
However, their ontological framework does not consider the idea of a protocol-driven conversation.
As a confirmation, in~\cite{Attardo17general} the author claims that
\begin{quote}
the term “narrative” strategy was a misnomer, as it might have given the impression that the GTVH was trying to handle narratological concerns, which are mostly beyond its scope.
\end{quote}
Interestingly, however, Attardo also notices that ``a three-step sequence [is] frequently used in jokes because it is the smallest number of repetitions necessary to set a pattern of expectations and breaking it''.

A three-step sequence was also identified in conversations by Glenn and Holt in~\cite{Glenn17conversation}.
The same structure was previously observed by Rozin et al. in~\cite{Rozin06aab} who called it the AAB pattern.
Noticeably, they even proved that the AAB pattern is statistically predominant in music and puns.
Although they never mention the idea of a protocol, several of their examples consist of jokes in the form of a conversation between two characters.
In our opinion, aporia protocols provide a more formal definition of the three-step structure that some authors intuitively identified. 

Finally, some authors already used artwork as case studies and working examples.
In particular, some of the examples presented here, as well as the general idea behind the paper structure, have been inspired by previous work of Luca Viganò~\cite{Vigano21moon}.

\section{Conclusion}
\label{sec:conclusion}

Wittgenstein's claim that ``A serious and good philosophical work could be written consisting entirely of jokes''~\cite{Malcolm62wittgenstein} finds a novel interpretation under the model presented in this paper.
As a matter of fact, assuming that ``A serious and good philosophical work'' amounts to a collection of arguments on the truth of philosophical constructs that the author wants to convey to the reader, she can rely on $\Sigma$-protocols to build PoK.
Since PoK and aporia protocols share the same structure, they can coexist in a single session.
This is something that satiric authors regularly do.
Even more, since aporia protocols can convey different emotions, Wittgenstein's statement also applies to, e.g., scary and angry works.
Examples show that the emotional message of artworks is mostly independent from the underlying communication protocol and, thus, it can be modified, e.g., as in parodies.

Although our proposal applies to many cases and contexts, there is no evidence that this model is universal and other protocols for transmitting emotions might exist as well.
However, since $\Sigma$-protocols are extremely succinct (only three messages), they might be a privileged structure, e.g., from an evolutionary perspective efficient communications might be an advantage.
In this respect, further research is needed to confirm the generality of aporia protocols.

Finally, thank to their algorithmic nature, aporia protocols can be used for better understanding intelligence in general.
For instance, we can use them for implementing new types of AI and for detecting complex communications in other beings.
The investigation of these, as well as other applications is left as future work.



\section*{Acknowledgment}
The author thanks Maria Luisa Catoni, Gustavo Cevolani, Daniele Fabbri, Giorgio Montanini and Luca Viganò for their insightful comments and for the useful discussions.

\bibliographystyle{plain}
\bibliography{biblio}

\end{document}